# Few-layer flakes of Molybdenum Disulphide produced by anodic arc discharge in pulsed mode

Carles Corbella[1,*], Sabine Portal[1], M.A.S.R. Saadi[1], Santiago D. Solares[1], Madhusudhan N. Kundrapu[2], Michael Keidar[1]

[1] Department of Mechanical & Aerospace Engineering, George Washington University, 800 22nd Street, Northwest, Washington, DC 20052, United States of America

[2] Tech-X Corporation, 5621 Arapahoe Ave. Suite A, Boulder, CO 80303, United States of America

* E-mail: ccorberoc@gwu.edu

## Abstract

Here, the synthesis of Molybdenum Disulphide ($MoS_2$) flakes by means of anodic atmospheric arc discharge is reported for the first time. The vertical electrode configuration consisted of a compound anode (hollow graphite anode filled with $MoS_2$ powder) and a solid graphite cathode placed just above of the compound anode. Arc processes were operated in pulsed mode to preferentially evaporate the powder component from the anode and to minimize Carbon ablation. Pulsed anodic arc discharges were conducted at 2 Hz and 10% duty cycle in 300 Torr of Helium with a peak current of 250-300 A and peak voltage of 35 V. A probe made of Tungsten wire was placed in the vicinity of the arc column to collect the evaporated material. The measured thickness profile was correlated to the particle flux distribution and it was fitted by a simple model of plasma expansion. During pulse phase, electron density was estimated around $5 \times 10^{22}$ $m^{-3}$ or higher, and ion current density was of the order of 10 $A/mm^2$. Morphology, structure and composition of the samples were characterized by Raman spectroscopy, atomic force microscopy (AFM), scanning electron microscopy (SEM), transmission electron microscopy (TEM), and x-ray diffraction (XRD). The study shows that pulsed arc discharge of the compound anode leads to moderate C deposition combined with $MoS_2$ deposition in the form of fragmented nanocrystals and few atomic monolayers of $MoS_2$. Such synthesis technique is promising to produce new 2D nanomaterials with tailored structure and functionality thanks to the flexibility of pulsed power.

# 1. Introduction

Nanomaterials are qualified for many applications in electronics, mechanics, energy, etc., thanks to their special physical properties.[1-4] The specific geometries of such materials are associated with particular crystalline structures and chemical functionalities.[5] Therefore, only a reduced subset of compounds fulfils the conditions to be synthesized in nanomaterial form. The three best known examples are constituted by materials with a natural tendency to forming atomic planes. Their in-plane atoms are arranged by strong covalent bonds which yield weakly interacting adjacent sheets by van der Waals forces, namely: *(1) Carbon nanostructures*, whose allotropes graphene, carbon nanotube and Buckminster fullerene are recognized worldwide as versatile materials with unique properties;[6,7] *(2) hexagonal Boron Nitride (h-BN)*, which has been synthesized e.g. in nanotube and nanosheet shapes, is promising for many applications as it is isostructural with graphene ('white graphene'),[8,9] and *(3) Molybdenum Disulphide (MoS$_2$)*, a transition metal dichalcogenide that forms a molybdenite crystal showing extraordinary electronic, tribological and biocompatible properties.[10-12] The cited materials can be obtained by means of different synthesis techniques, such as chemical or mechanical exfoliation of bulk material,[13] and plasma processing of a gaseous precursor.[14]

Anodic arc discharges constitute very efficient plasma sources of nanomaterials, as initially proven by the seminal work by Iijima on the production of carbon nanotubes from plasma ablation of a graphite anode.[6] The high gas temperatures reached in the carbon arc column provide appropriate conditions for synthesis of carbon crystalline nanostructures. Moreover, the control of the plasma parameters through setting electrical parameters of arc current and arc voltage, working pressure and electrodes geometry, makes arc discharge an attractive laboratory technique to investigate the basic processes involved in nanomaterial synthesis.[15] This same method has proven to be effective to grow graphene flakes, i.e. few layers or single layers of carbon atom crystals.[16,17] Further efforts towards the production of 1D and 2D materials include the research on atmospheric arc plasmas aimed to h-BN growth. Nanotubes of h-BN were grown by arcing a h-BN pressed rod inserted into a hollow Tungsten electrode. The W electrode was necessary to hold the discharge because h-BN is insulating.[18] In another example, boron ingots with superficial metallic inclusions have been used for h-BN production in arc processes. The design of such compound anodes with metallic inclusions, whose mission is to enhance arc current, has permitted the successful synthesis of h-BN nanotubes by reactive arc discharge in nitrogen atmosphere.[19]

Concerning $MoS_2$ production, although plasma synthesis of 2D flakes and fullerene-like nanoparticles of $MoS_2$ has already been addressed,[10,20] synthesis recipes by means of anodic arc discharge in gas phase have not been reported so far. On one hand, the rapid melting of the molybdenum anode in reactive arc synthesis of $MoS_2$ constitutes an important issue in process control and stability.[21] This issue has been confirmed in a recent study reporting plasma parameters during Mo anode erosion by atmospheric arc discharge.[22] On the other hand, Sulphur-containing reactive gases like $H_2S$ are routinely used to deposit $MoS_2$ in reactive processes despite their toxicity.[20,23]. The drawbacks mentioned above motivate the search of alternate methods of $MoS_2$ synthesis, which avoid the use of toxic reagents and offer acceptable repeatability and control over plasma parameters. One strategy to avoid handling of Sulphur-containing gases consists in arcing a $MoS_2$ powder precursor filling a hollow anode made of a refractory and conducting material such as graphite. The use of such container with high electrical conductivity is indispensable to strike and maintain the arc discharge, which otherwise would not be possible because $MoS_2$ is a semiconductor, and thus, would prevent arc formation. Furthermore, the shortcoming of early Mo melting suggests the need for designing novel arc discharge recipes. A proposed solution might be energizing the anode through a periodic signal waveform, such as a square-wave modulated pulsed current. This approach has already shown positive results in carbon arc discharge research.[24]

The present study has two objectives: **(1)** to prove that anodic arc discharge is a viable technique to deposit few-layer stacks of $MoS_2$, and **(2)** to propose a non-reactive route of $MoS_2$ synthesis based on pulsed anodic arc discharge from $MoS_2$ powder precursor. The second objective is motivated by the special plasma properties of pulsed arcs, which have been introduced recently for carbon nanomaterials synthesis application. In the article by Corbella et al., pulsed plasmas in arc discharges applied to a solid graphite anode show a series of advantages, including significant dilution of amorphous carbon macroparticles, better control of arc processes, and production of nanomaterials in more stable discharges.[24] An interesting feature of pulsed arc discharges is its repeatability through the regular localization of the arc core at the intermediate position between anode and cathode. This effect makes the method adequate for preferential evaporation of the central part of the anode, in contrast with the irregular and chaotic ablation of anode characteristic of steady DC arc discharges.[25,26] Hence, pulsed anodic arc discharge is the selected plasma method to enhance evaporation of the $MoS_2$ powder filler while simultaneous erosion of the graphite container is minimized. In this article, the parameters of the pulse signal have been adopted according to past studies,[24] and are applied for the deposition of $MoS_2$ onto a wire substrate (collecting probe). Plasma parameters like particle flux distribution and electron density have been quantified. The fabrication of $MoS_2$ flakes in this arc process has been discussed by means of atomic force microscopy (AFM) and Raman spectroscopy measurements on the deposited sample. Finally, the morphological, structural and chemical properties of the sample are characterized by electron microscopy and X-ray analysis.

## 2. Experimental Details

*2.1. Experimental Setup*

The anodic arc discharges were performed using an experimental setup described in detail elsewhere.[24,27] The arc discharges were held inside a cylindrical arc plasma chamber of 4500 cm$^3$, which was pumped down by a mechanical pump, down to a background pressure lower than 0.1 Torr. The processes were carried out at a working pressure of 300 Torr in a He atmosphere (purity: 99.995%). The electrodes were vertically aligned inside the vacuum vessel. In this study, a compound anode was used to evaporate $MoS_2$. Such anode consisted of $MoS_2$ powder inserted into a graphite container. A hollow graphite anode, with inner diameter of 3 mm and outer diameter of 5 mm, acted as precursor container. The anode was filled with $MoS_2$ powder (purity: 99%) with grains of 1 μm average size, acquired from Stanford Advanced Materials. The compound anode was mobile and placed at the lower position, and a fixed solid graphite cathode of 10 mm in diameter was located above the anode. A Miller SS300 DC power supply was used to ignite and support the discharge. The arc discharge was ignited by energizing the electrodes, initially brought into electrical contact, and separating them afterwards. For this, the position of the lower anode was controlled by means of a linear drive. The DC power supply was remotely operated in pulsed mode by means of a waveform generator, which supplied rectangular pulse signals of 5 V amplitude (yielding up to 300 A of peak current) at 2 Hz frequency and 10% duty cycle. These parameters provided maximal stability and lifetime of the arc discharge. In addition, DC arc processes held at 60 A of arc current were carried out for comparison purposes.

The arc discharge processes were characterized by studying the V-I characteristics of the discharge.[22,28] The electrical circuit is schematized in Fig. 1a. Moreover, a collecting probe consisting of a Tungsten wire of 0.5 mm in diameter and approximately 30 mm long was installed horizontally at a distance of 6-7 mm from the arc core position to measure the particle flux distribution in the vicinity of the discharge. The probe was not placed closer to the arc column to avoid melting issues.[27] Fig. 1b and Fig. 1c show the respective top view and lateral view of the setup with the collecting probe, which was placed next to the plasma gap between compound anode and solid cathode. After arc process was completed, the probe was left inside the chamber for at least 20 min to cool down.

*2.2. Characterization techniques*

The bonding structure of the samples was characterized by Raman spectroscopy using a Horiba LabRAM HR operated at a wavelength of 532 nm. The analyzed spot sizes were of the order of a few microns. Surface topography was explored by AFM using an Asylum Research instrument (MFP-3D) operating in tapping mode. Imaging was conducted in the attractive regime to avoid surface damage and contamination of the sample by the AFM probe. The sample was profiled with a Si cantilever (Asylum Research, AC200TS, with resonance frequency $f \approx 151$ kHz and force constant $k \approx 6$ nN/nm). The scan rates used for the images reported in Fig. 6a and 6b were 1 Hz and 0.5 Hz, respectively. The morphological characterization was carried out by means of scanning electron microscopes (SEM) Tescan XEIA FEG SEM and Tescan GAIA FEG SEM with accelerating voltage ranging between 5 kV and 25 kV. Both SEMs were equipped with energy dispersive X-ray spectrometry (EDS) setups, which were used to measure the chemical composition of the samples. High-resolution transmission electron microscopy (HRTEM) imaging and structural study by selected area electron diffraction (SAED) were performed with a JEOL JEM 2100 LaB6 TEM. Finally, crystallographic phases of the samples were analyzed by powder X-ray diffraction (XRD) by means of a Bruker C2 Discover system with 2D detector, which emits Cu $K_\alpha$ radiation (1.54056 Å) and is equipped with a Vantec500 detector especially suited for obtaining quality patterns from small samples. This instrument provides parallel beam geometry to probe small areas or special shaped samples.

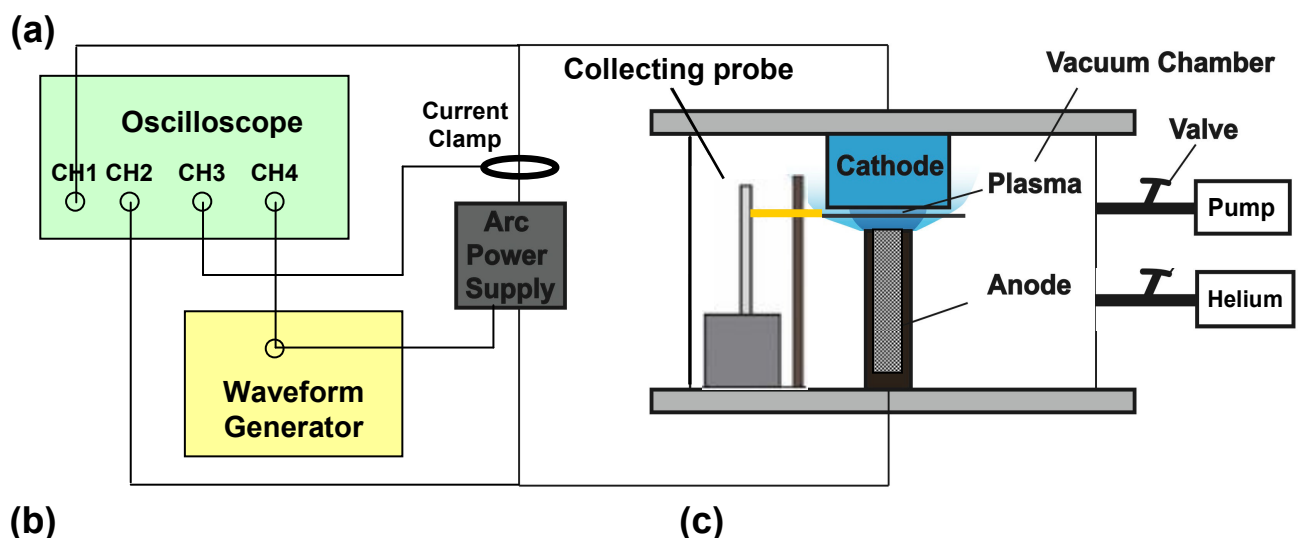

**FIG 1:** (a) Experimental setup of the arc plasma chamber together with electrical measurement system (adapted from Corbella et al. [24]). Compound electrode of hollow graphite anode filled with $MoS_2$ powder ($MoS_2$@graphite) is installed in the plasma arc chamber. (b) Top view and (c) lateral view of the collecting probe. The collecting probe, which consists of a W wire ($\phi$=0.5 mm) inserted into a ceramic tube, is placed at a distance of 6-7 mm from the anode axis.

## 3. Results and Discussion

*3.1. Deposition process*

*3.1.1. Plasma parameters in pulsed arc discharges*

Fig. 2a schematizes the typical ablation modes and light emission patterns observed in steady DC arc discharge (60 A) and pulsed arc discharge (250-300 A) in He atmosphere. Briefly, pulsed arc discharge is developed preferentially in the intermediate region of the plasma gap, while DC arc discharge takes place at the surrounding region of the anode.[24] In particular, this type of discharge promotes ablation of the edges of the anode. Fig. 2a illustrates also in more detail the application of pulsed arc discharge on a compound anode like the one used in this study. The periodic ignition of plasma arc generates a hot gas region that interacts with the powder, thereby evaporating the $MoS_2$ powder with minimal erosion of the graphite container. Once the $MoS_2$ grains are within the hot gas atmosphere, they undergo fragmentation and molecular dissociation. Then, the ionized or neutral reaction products are transported through the plasma and gas phases until they impinge onto a deposition surface. This sketch constitutes the scenario of pulsed arc discharges on the powder-containing anode, whose electrical parameters are evaluated and discussed below.

Fig. 2b shows the periodic arc voltage and arc current waveforms recorded during arc evaporation of the compound anode in pulsed mode at 2 Hz and 10% duty cycle. The representation of V-I characteristics (Fig. 2c) shows the cyclic behaviour typical of pulsed arc discharges generated with the DC power supply mentioned above. Also, the fact that V-I curves float above the threshold voltage is a signature of arc plasma ignition, as investigated in a previous study.[22] Measured values of peak current (≈ 250-300 A) and of peak voltage (≈ 35 V) are contrasted with the lower current (225 A) and higher voltage (50 V) recorded in carbon pulsed discharge. These values suggest a plasma conductivity of a factor ≈ 2 compared to carbon arc experiments, as inferred from the respective ratios of current and voltage. With this comparison, one can estimate plasma parameters since electron density is proportional to plasma electrical conductivity.[22,29] From the signal values measured here, a lower estimate of plasma density should be roughly twice the plasma density of carbon discharge (≈ $3 \times 10^{22}$ m$^{-3}$), resulting in around $5 \times 10^{22}$ m$^{-3}$. A circumstance supporting enhancement of plasma density is that the ionization energy of Mo atom, 7.1 eV, is lower than the ionization energy of C atom, 11.3 eV. Higher electron density might be also caused by the topography of the hollow anode: indeed, the electrode top surface might exhibit a stronger electron field emission stimulated by sharp edge effects. The ion current at the cathode surface can be determined by assuming the Bohm condition at a collisionless cathode sheath-edge:[30,31]

$$I_i = \pi r_A^2 0.6 e n_e \left( \frac{\kappa T_e}{M_i} \right)^{1/2} \qquad (1)$$

where $r_A$ is the anode radius, $e$ is the elementary charge, $n_e$ is the electron density, $\kappa$ is Boltzmann's constant, and $M_i$ is the ion mass. The ion beam area at the cathode surface has been taken as the full section area of the compound anode ($r_A$ =2.5 mm) because the deposited material onto the cathode shows a circular profile comparable to the anode circular section. With such assumption, and by taking an electron temperature of ≈ 1 eV, the ion current density has been calculated as function of the dominant ion in the arc discharge. The dominant ion is determined by the plasma chemical reactions, which have not been modelled in this article. For the sake of simplicity, here we consider two extreme cases in which dominant ions in plasma are $C^+$ ions ($M_i$=12 u) and $Mo^+$ ions ($M_i$=95.9 u). The obtained values range between 270 A and 100 A out of a total arc current of around 275 A, respectively. The respective ion current densities are 15 A/mm$^2$ ($C^+$: 8.5×10$^{21}$ cm$^{-2}$s$^{-1}$) and 5 A/mm$^2$ ($Mo^+$: 3.0×10$^{21}$ cm$^{-2}$s$^{-1}$). In arc experiments, He atoms are considered neutral since Helium shows the highest ionization potential (24.6 eV) of all species in the gas mixture. Exclusively atomic ions have been assumed to contribute to ion current since the Mo-S bond energy in $MoS_2$ is only 2.56 eV.[32] This energy is much lower than the atomic ionization potentials, thereby suggesting a high dissociation degree of $MoS_2$ molecules in plasma. A more accurate insight into pulsed plasma parameters would require the application of diagnostics like time-resolved optical emission spectroscopy (OES) or electrostatic Langmuir probe, which are planned in a future work aimed at pulsed plasma characterization.

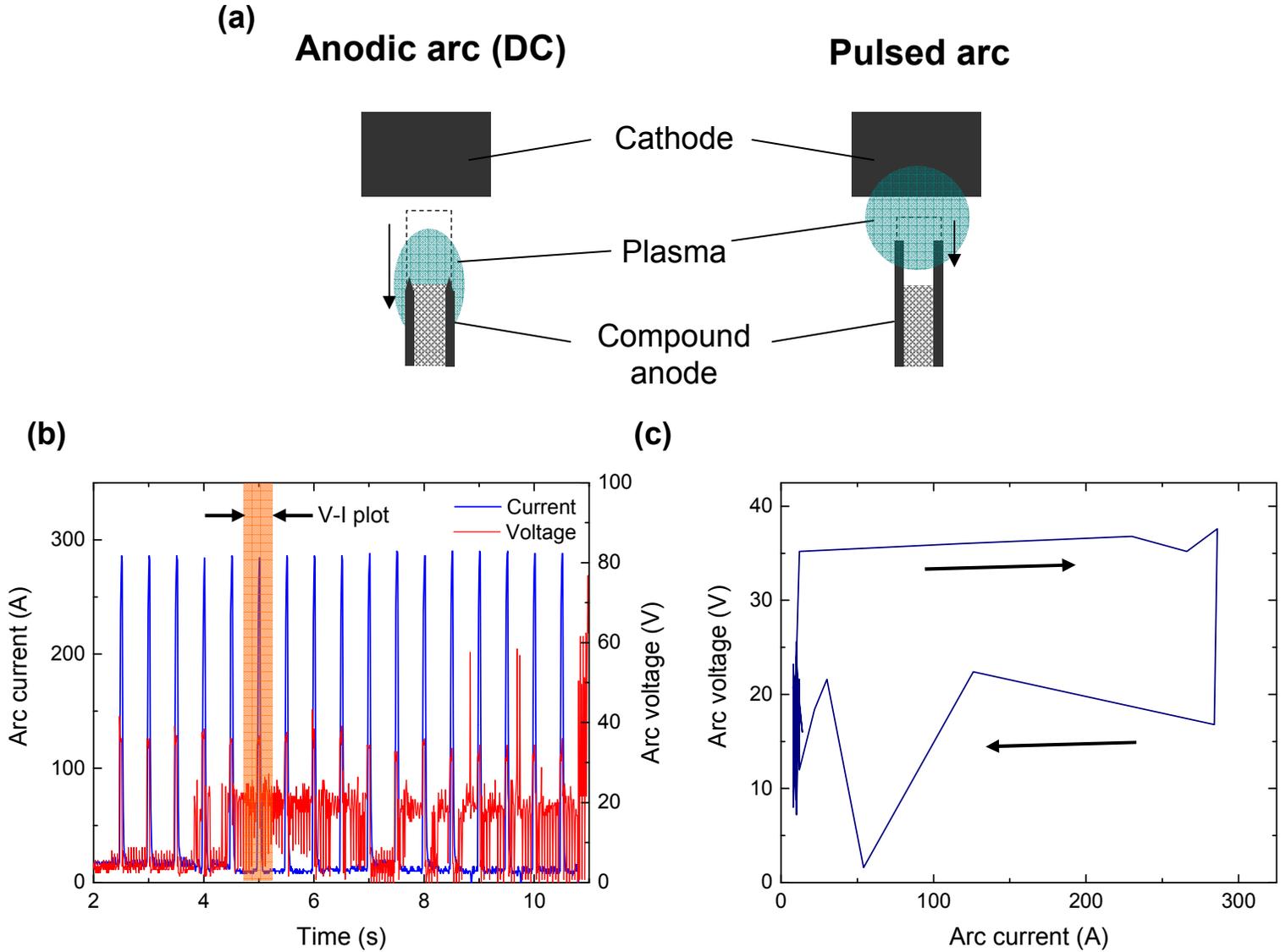

**FIG 2:** (a) Sketches (lateral views) comparing the light emission patterns and ablation modes during steady DC arc discharge and pulsed arc discharges on solid anode. The pulsed arc core is located at the intermediate position between anode and cathode (preferential evaporation of anode centre), whereas steady DC arc discharges provide a glow region dominating around the anode (preferential ablation of anode container). Therefore, the use of pulsed power is adequate to ablate preferentially the near-axis region of the anode, which leads to dominant evaporation of filling powder in compound anode. (b) Temporal evolutions of arc current (blue) and arc voltage (red), and (c) the representation of the shadowed pulse as V-I characteristics. The arrows indicate the direction of time within the cycle.

*3.1.2. Particle flux distribution*

Fig. 3 shows an image of the probe with deposited layer and two distributions of incident particle flux onto the collecting probe. They correspond to pulsed anodic arc discharges (2 Hz, 10% duty cycle) using, first, a solid graphite anode, and after that, a compound anode containing $MoS_2$ precursor powder. The deposition study of Carbon has been performed to test this method. The probe had to be placed at a minimal distance from the arc column to prevent excessive heat transfer from the arc discharge to the probe. An arc-probe distance between 6 and 7 mm from the arc centre was chosen. This selection is based on earlier measurements performed with a fast-moving probe within the same arc plasma chamber.[27] There, simulations of gas temperature profiles in steady DC arcs provided limit values of around 3500 K at distances larger than 5 mm from the arc core (W melting point: 3695 K). Moreover, probe melting was not observed in regions at 3 mm or further from arc centre for any studied exposure time (between 10 and 60 ms). In our study, the probe is at a fixed location and pulse duration is 50 ms as defined by the duty cycle (10% of 500 ms period). Since arc current saturation is not established for such a short pulse time,[22,24] the effective exposure time of the probe to the pulsed discharge is less than 50 ms in each period. This fact, in combination with the safe distance to the arc core, leads to conclude that the W probe was not significantly damaged during the experiments.

The particle flux distribution in each experiment has been calculated by measuring the layer thickness at different positions $x$ of the probe (for $MoS_2$: ≈ 25 μm at the middle and ≈ 5 μm at the extremes for a 8 s-process of $MoS_2$ evaporation) and by converting the deposition rate, $R(x)$, into particle flux density projected in the X direction, $J_x(x)$. For this, the following equation was used:

$$J_x(x) = \frac{\rho R(x)}{M} \qquad (2)$$

where ρ is the material mass density and $M$ is the molecular mass of the impinging species from the plasma. Therefore, assuming a radial particle flux emerging punctually from the arc core, the particle flux at a distance $r$ from the arc core can be calculated as $J(r)=J_x(x)/\cos\theta$, being θ the angle subtended by the normal direction from probe to arc core and the segment comprised between arc core and position $x$ at the probe (see inset in Fig. 3a). The mass densities for Carbon and MoS$_2$ evaporation processes were approximated to 2 g/cm$^3$ and 5 g/cm$^3$, respectively. The atomic mass considered in graphite ablation is the one of Carbon, 12 u, and the elementary mass considered for MoS$_2$ evaporation is the corresponding molecular mass, which is 160 u. This description accounts for a simple model since the flux incident onto the probe surface consists in reality of a combination of atoms and molecules formed by Mo and S elements produced by the plasma chemistry reactions. Influx of C species is also expected due to graphite container ablation. The layer chemical composition is analyzed in section 3.2.

The dependencies displayed in the plots confirm the assumption of spherical propagation of particles from the arc core since the flux density obeys a law ~$1/r^2$. The particle flux at the anode surface (far left data point), $\phi_{anode}=E/(M·S)$, is calculated from the anode cross section (3 mm in diameter), $S$; the atomic masses of the evaporated elements, $M$; and the evaporation rate of each experiment, $E$. The parameter $E$ was obtained by dividing the removed mass (measured with a microbalance) by the process duration. In the case of solid graphite arc discharge, the average ablation rate was 1 mg/s. In the case of arcing of the MoS$_2$@graphite compound anode, the total evaporation rate was estimated to be around 75 mg/s, whose particle flux density is only assigned to MoS$_2$ species and is roughly a factor 5 higher than carbon flux density. The underestimation in the calculated central flux in MoS$_2$ evaporation is probably due to a low sticking coefficient of MoS$_2$ species onto the collecting probe and/or growing layer (see Fig. 3c). Further probe experiments were performed using steady DC arc power at 60 A applied to the compound anode (not shown here). It is worth noting that, in these experiments, no radial distribution of plasma expansion of the kind ~$1/r^2$ could be inferred. Instead, the layer grown on the collecting probe showed an irregular deposition profile. This result is related to the irregular and instable structure of DC arc discharges, compared with the more stable and easy-to-control pulsed arc discharges.

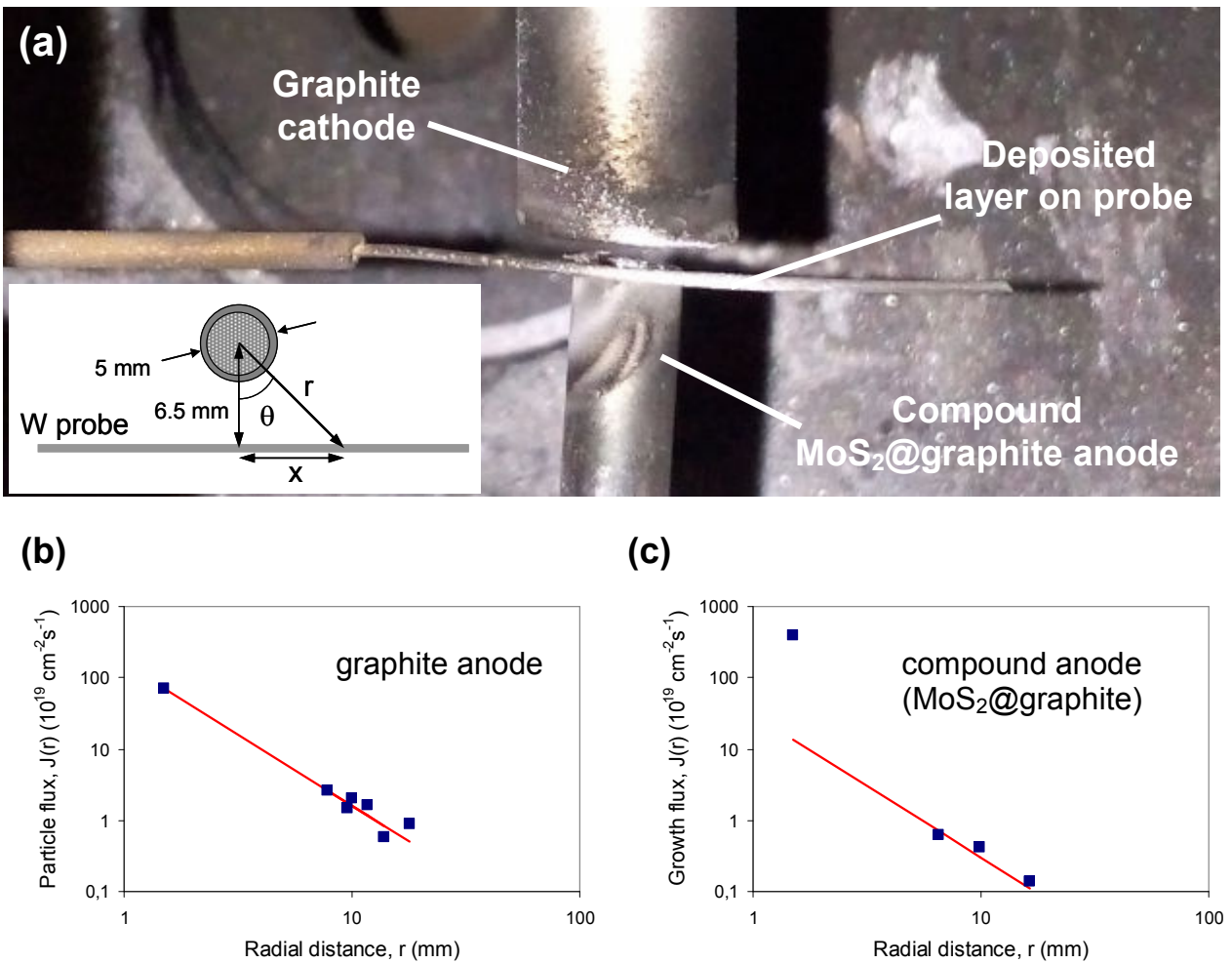

**FIG 3:** (a) Thickness profile along the collecting probe after MoS$_2$ pulsed arc evaporation process. The particle flux is calculated in different probe zones and represented as a function of the radial distance to the arc column axis. Inset: schematic top view of the probe and compound anode with definition of the model parameters. First, the model was tested for pulsed arc graphite ablation profile (b), and afterwards the model was applied to pulsed arc $MoS_2$ evaporation profile (only to the three data points from the probe) (c). A model (red line) fitted to the experimental data shows a $\sim 1/r^2$ dependence of the particle flux for both solid graphite ablation and compound $MoS_2$@graphite evaporation.

*3.2. Sample characterization*

*3.2.1. Bonding structure (Raman)*

The layer deposited onto the W probe was analysed by Raman spectroscopy. Fig. 4 shows the explored regions together with the corresponding Raman shift spectra. The middle region of the probe was mostly covered with grains from $MoS_2$ precursor and their fragments. The two characteristic Raman bands of $MoS_2$, which correspond to the phonon modes $E^1_{2g}$ ($\approx$ 385 cm$^{-1}$) and $A_{1g}$ ($\approx$ 410 cm$^{-1}$),[33] show a frequency shift separation of 25 cm$^{-1}$. According to literature, this inter-band distance corresponds to $MoS_2$ in bulk form.[12] Bands with identical separation are observed along the probe. There, fragments of $MoS_2$ grains from the powder precursor coexist with an amorphous carbon phase, as proven by the measurement of broad D ($\approx$ 1350 cm$^{-1}$) and G ($\approx$ 1550 cm$^{-1}$) Raman bands characteristic of Carbon.[34] The Raman spectra of DC arc-coated probes (not shown here) display majority of carbon deposition. The dominant C concentration is correlated to preferential graphite anode ablation in DC processes.

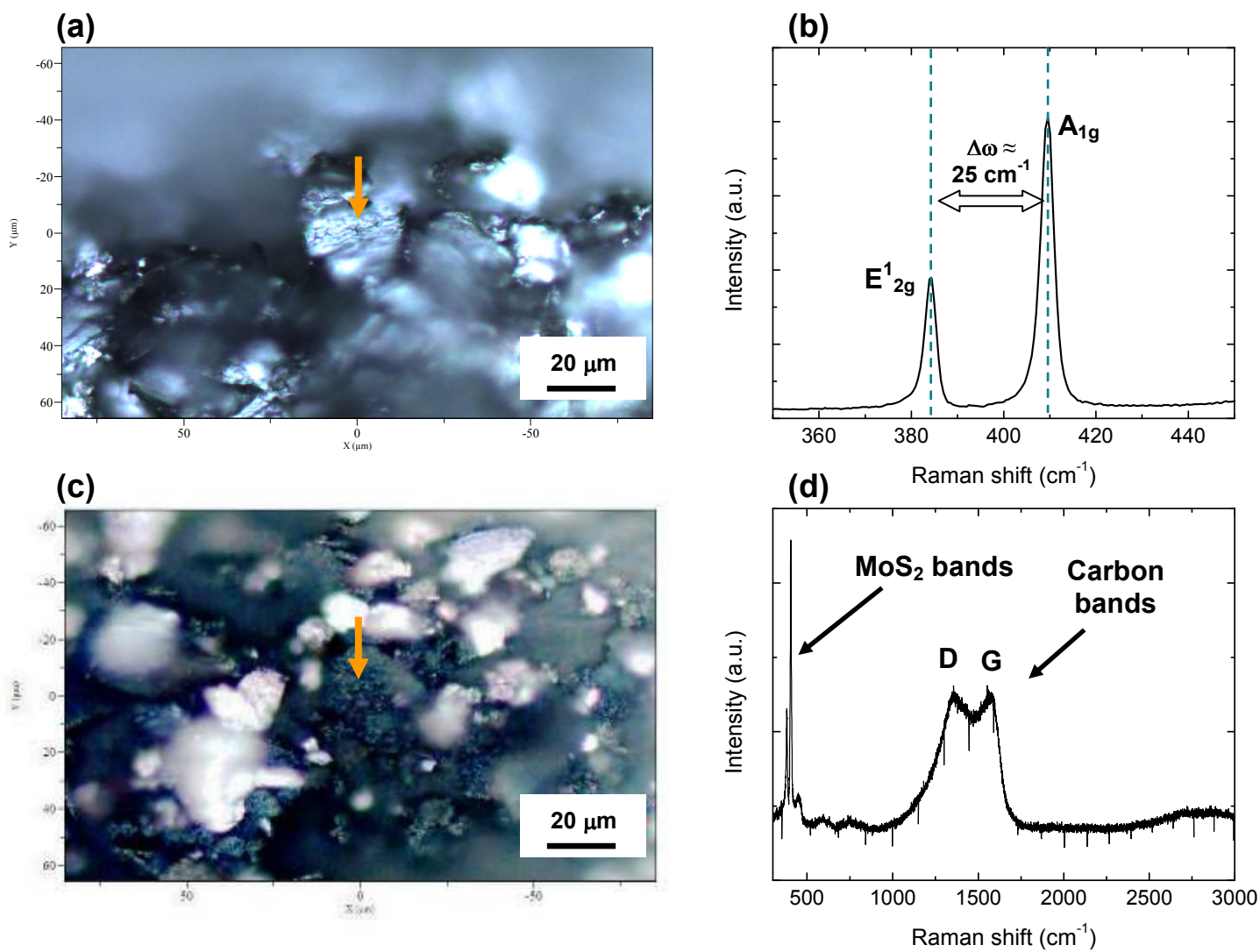

**FIG 4:** (a), (c) Optical microscopy images and (b), (d) Raman spectra of $MoS_2$ layers deposited on the collecting probe. The analyzed regions are a few microns size (see arrows). The two phonon modes $E^1_{2g}$ and $A_{1g}$ are indicated. The structure changes with the distance to the arc axis (a,b: middle region; c,d: extreme region). Raman shift peak separation at a $MoS_2$ grain facet is 25 cm$^{-1}$, which is characteristic of bulk or many-layers $MoS_2$. Carbon and $MoS_2$ appear combined in certain conditions: the deposited grains dominating in the middle probe region (probably projected from the plasma expansion) show only $MoS_2$ signals (b), whereas both materials coexist combined at the background surface in the extreme region (d).

Fig. 5 shows images of the probe surface in regions where the grown layer was very thin after a gentle scratch of the sample surface. The probed regions after mechanical thinning showed small, micron-size islands, which are termed Region 1 (Fig. 5a) and Region 2 (Fig. 5b). The measured $MoS_2$-characteristic Raman bands on such regions were located at positions like those shown in Fig. 4, but they appeared closer to each other, with a separation ranging between 23 cm$^{-1}$ and 24 cm$^{-1}$ (Figs. 5d and f). Moreover, the concentration of Carbon is negligible in such regions as concluded from the lack of Carbon Raman shift signals (Fig. 5c). The inter-band distance measured here corresponds to $MoS_2$ in layered form, comprising between 3 and 5 elementary layers.[12] This result is an evidence of few-layer flakes of $MoS_2$ deposition on the probe surface. However, it is not clear neither where the growth process preferentially takes place nor what the growth mechanism is. Probably, elementary species from the arc plasma, such as Mo atoms, S atoms and molecular species impinge onto Tungsten surface, and nucleate and generate 2D layers on surface topographies favourable for this type of growth, such as terraces. Nevertheless, the formation of $MoS_2$ nanoparticles in gas phase cannot be discarded and should be explored further with real-time plasma diagnostic techniques. The presence of $MoS_2$ flakes is supported by AFM analysis, which is discussed in the following subsection.

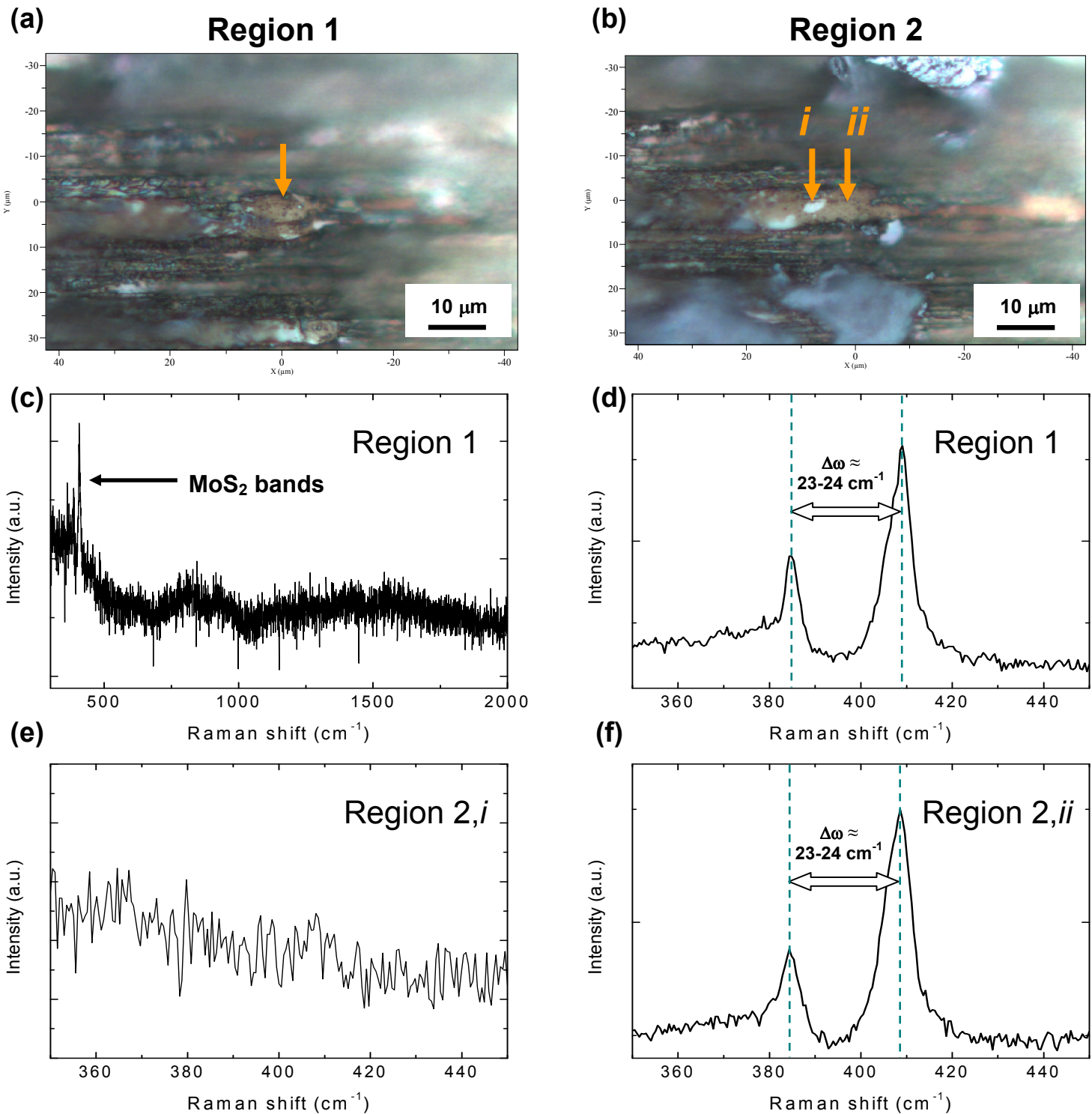

**FIG 5:** Optical images of MoS$_2$ islands on the collecting probe. They are located in thin layer regions and have been labelled as Region 1 (a) and Region 2 (b). Few-layer flakes of MoS$_2$ are attached to the W surface of the collecting probe. (c) Raman spectrum of Region 1 showing MoS$_2$ bands and a very small Carbon signal. (d) The separation between the characteristic MoS$_2$ peaks is as low as 23 cm$^{-1}$, which corresponds to few-layer

structures of MoS$_2$ (<5 layers). (e), (f) Raman spectra of zones *i* and *ii* of Region 2, respectively, which show optical contrast. The lack of MoS$_2$ signal in zone *i* suggests that there is a hole in the island (the substrate is exposed).

*3.2.2. Identification of flakes (AFM)*

The AFM images in Fig. 6 (flattened using a second order polynomial fit) show a cluster of few-layer flakes in the material deposited onto the collecting probe. The explored area coincides with the sample region where the Raman signal evidenced few-layer flakes of MoS$_2$ (Fig. 5). The measured step between neighbouring facets is of the order of 1 nm. Since the thickness of a MoS$_2$ monolayer is ≈ 0.65 nm,[12] it is plausible to assume that the depicted nanosheets consist of elementary 2D layers of MoS$_2$ located at the near-substrate region. Such flakes are exclusively attributed to MoS$_2$ compound and the possibility of being composed of any other material is discarded as judged from the Raman analysis (Fig. 5). The observation of the nanosheet cluster, which is composed of randomly distributed MoS$_2$ nanometric monolayers, is consistent with the prediction of few monolayers of MoS$_2$ as it is inferred from Raman measurements. The correlation between AFM and Raman measurements can be explained by the probing area of the laser beam used to obtain the Raman spectra. Indeed, the laser spot of the Raman spectrometer has a few microns diameter. The laser spot size is thus extended over an area that comprises several MoS$_2$ monolayers according to the AFM image. In this way, the contribution of each individual monolayer sums up to the Raman output (Fig. 5) which suggests an average number of 4 monolayers or less.

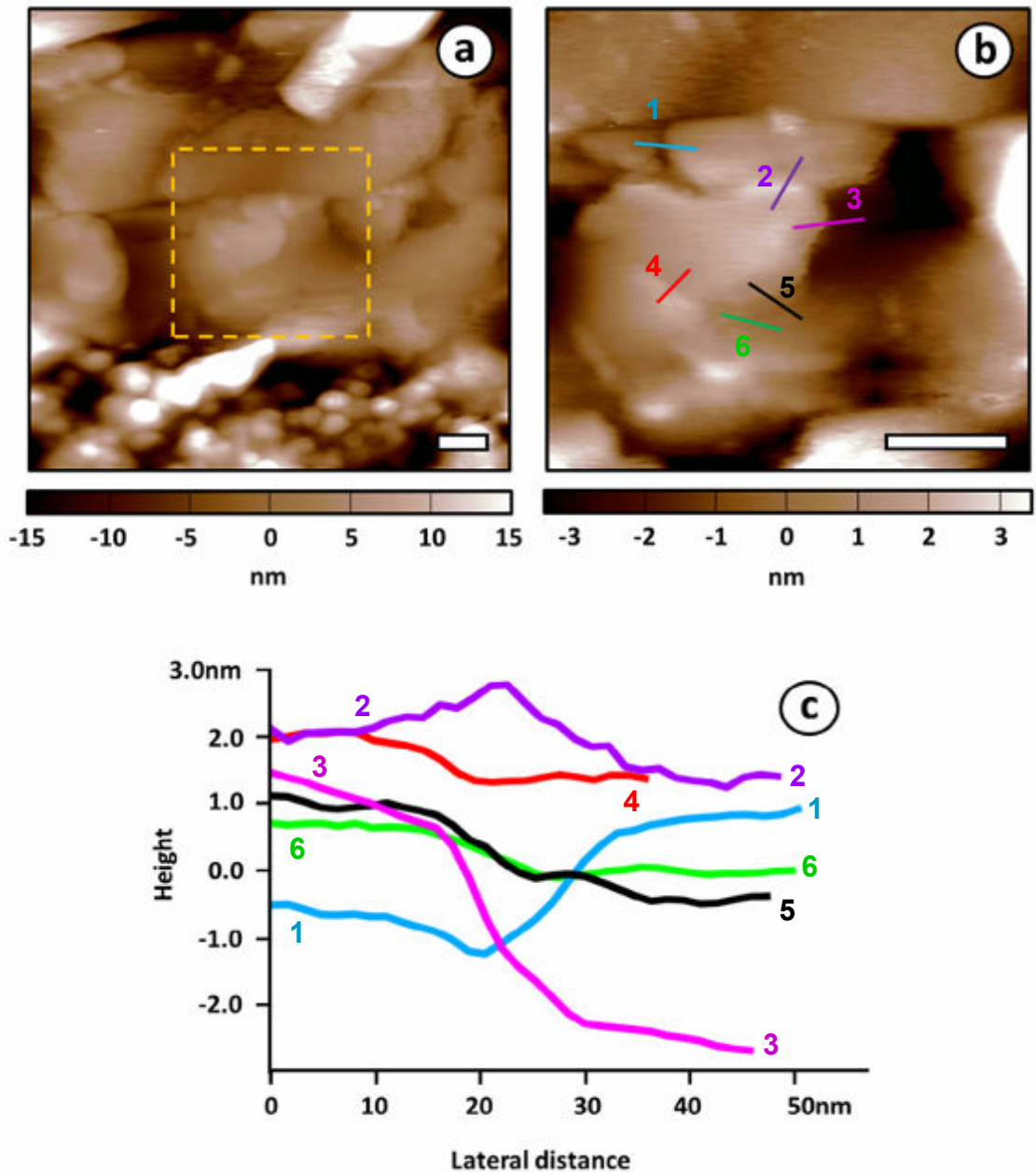

**FIG 6:** (a) AFM image of a cluster of 2D nano-flakes localized at the near-substrate region of the collecting probe. (b) Zoom in of the region delimited with a square frame in (a). (c) Height profiles of the scan lines marked in (b). The steps measured between neighbouring flakes (2-purple, 4-red and 6-green lines) show heights of around 1 nm, which match with the approximate thickness of one $MoS_2$ monolayer (≈ 0.65 nm). Scale bars in both images (Fig. 6a and 6b) correspond to 100 nm.

*3.2.3. Surface morphology and composition (SEM, EDS)*

Figs. 7a and 7b show SEM micrographs of the $MoS_2$ powder used as deposition precursor. The grains have a nominal average size of 1 μm and appear forming agglomerates that can achieve around 50 μm in size. Fig. 7c shows a portion of the W probe (middle region) which is covered by the deposited layer of $MoS_2$. A magnification of this region (Fig. 7d) evidences deposition of chipped fragments of the original $MoS_2$ grains and agglomerates. The presence of smaller-sized grains spread on the probe surface suggests that $MoS_2$ grains expelled from the compound anode during arc discharge undergo fragmentation due to the high temperatures achieved in gas phase. Analysis of the material composition and its implications in the experiment are discussed next.

Fig. 8 shows SEM images of the middle region of the probe together with EDS spectra. The probed regions, which are delimited with rectangles, correspond to a $MoS_2$ chip (Figs. 8a and 8c) and to the surface background (Figs. 8b and 8d), respectively. The insets in the EDS plots provide atomic concentrations of the dominant elements, namely Carbon, Oxygen, Molybdenum and Sulphur. Tungsten is always detected because the electron beam interaction plume reaches the W probe surface. Further elements (Al, F) might be impurities coming from powder and substrate material. Figs. 8a and 8c show that Mo and S are detected following the stoichiometry of the $MoS_2$ molecule (1:2) in the chip region. The Carbon signal might come from two sources: (i) Carbon deposited during arc discharge, and (ii) adventitious carbon adsorbed from exposure to the atmosphere. The Oxygen signal could be generated by O atoms adsorbed on the layer surface due to atmospheric oxidation of Mo. However, the measured low concentration of O atoms and its relatively large uncertainty (low intensity peak) may account for a weak surface oxidation on the $MoS_2$ chip.

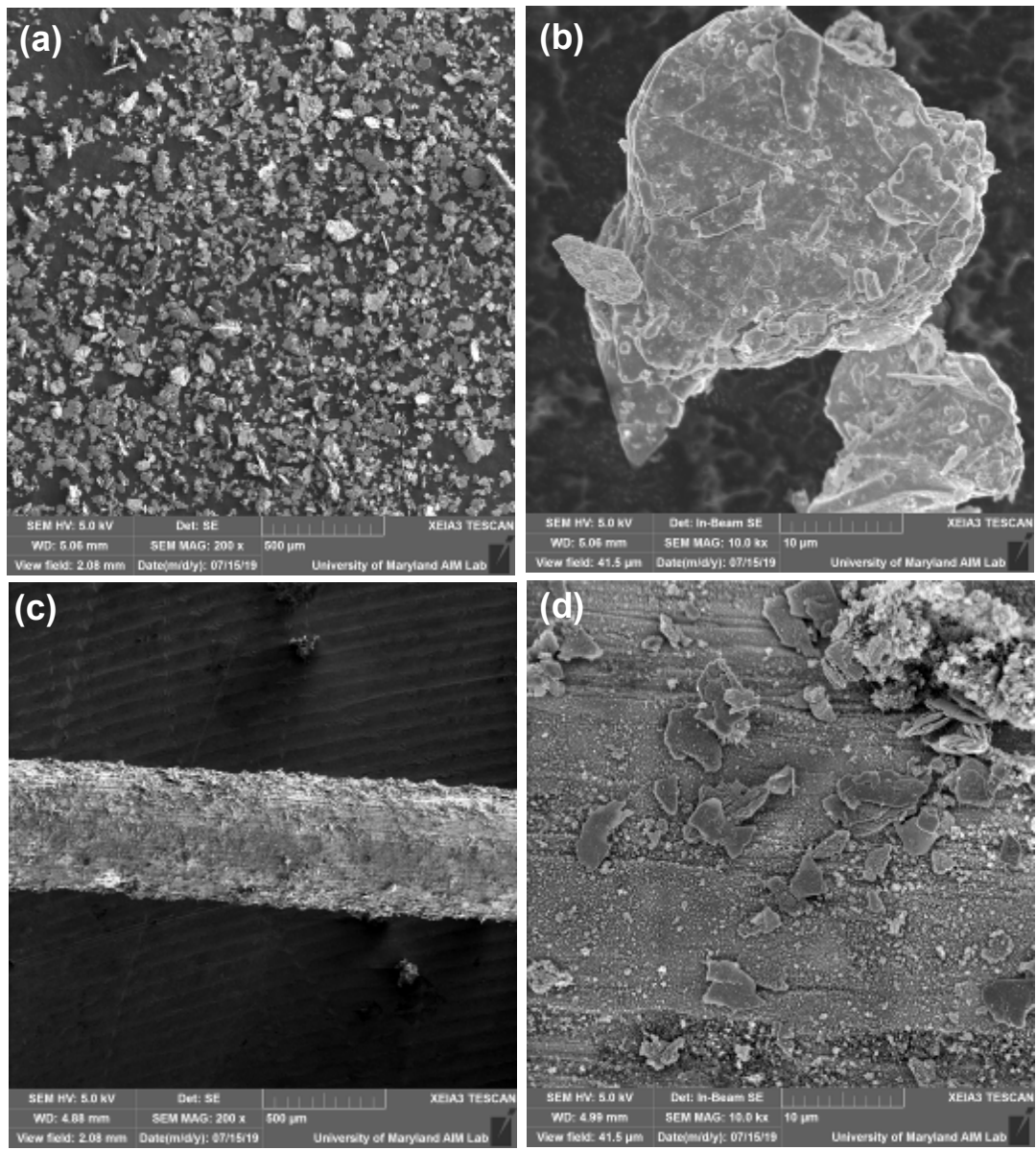

**FIG 7:** Top view SEM images of (a), (b) MoS$_2$ precursor powder (grains and aggregates) and (c), (d) MoS$_2$ deposition onto the middle region of the collecting probe. The morphology of the deposited material consists of fragments of the precursor powder distributed over a granulated background. The chemical composition measured by EDS is shown in Fig. 8.

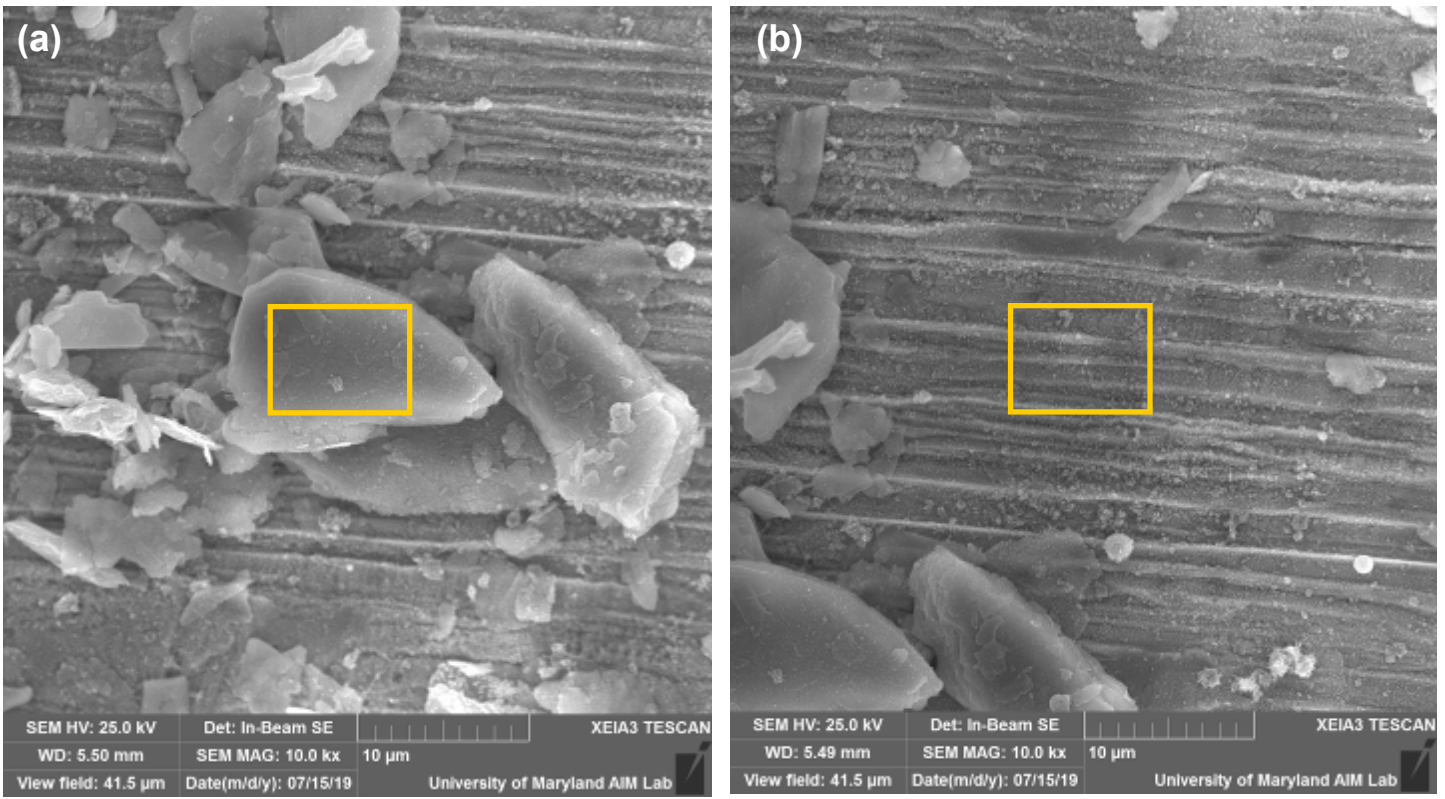
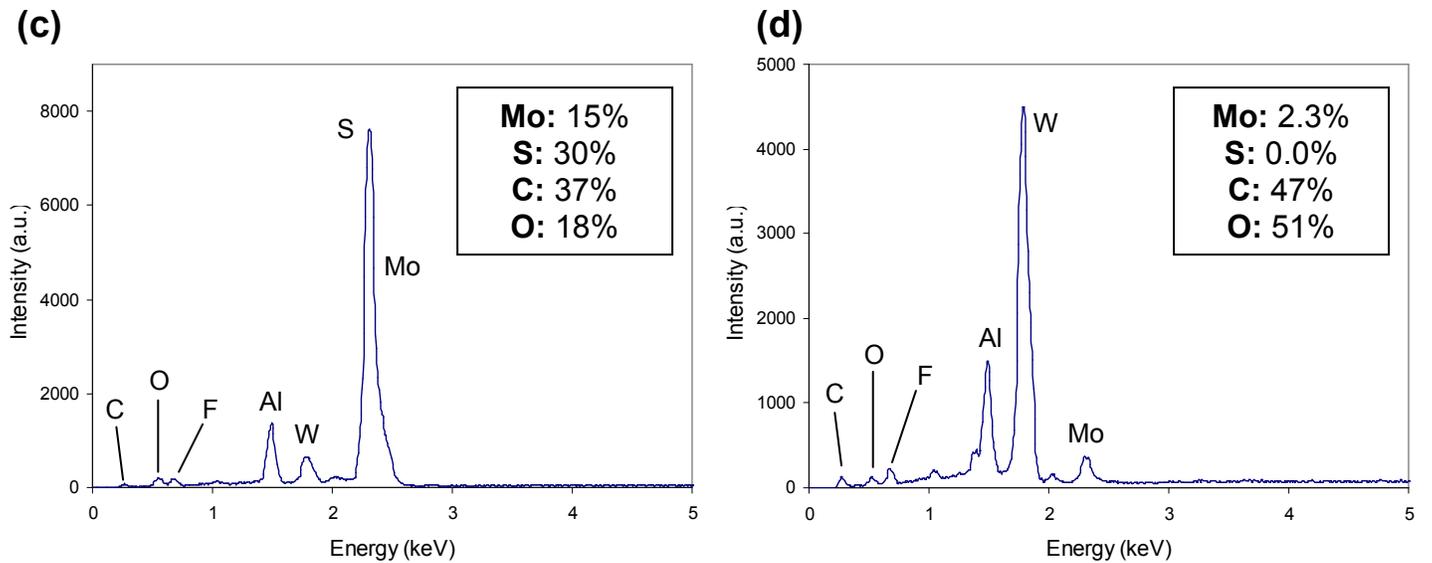

**FIG 8:** (a), (b) Top view SEM images of $MoS_2$ deposition at the middle of the collecting probe. (c), (d) Respective EDS plots corresponding to two different areas of the probe after deposition. Insets: atomic concentrations of the elements of interest. The grains (a) show stoichiometric $MoS_2$ composition, whereas the background surface (b) contains C, Mo and a large amount of Oxygen instead of Sulphur. Martincova et al. [35] suggested that O atoms might substitute S atoms in $MoS_2$ atomic monolayers in the atmosphere, creating $MoO_3$ molecules. The S atoms from near-substrate $MoS_2$ few layers may be undetected due their very small concentration compared to C and Mo amounts.

Figs. 8b and 8d show that the background surface of the sample is mainly composed by C and O atoms, while there is a very low atomic concentration of Mo. S atoms have not been detected. It is reasonable to assume that S atoms are only located at the interface between probe surface and deposited material in the form of few-layer flakes of $MoS_2$, fact that would agree with the absence of S signal in the EDS spectrum. The Mo detection without S might be explained by the following hypothesis. It has been reported that $MoS_2$ layers in presence of Oxygen undergo substitutional reactions at their edges, where S atoms are displaced by O atoms resulting in $MoO_3$ molecule formation.[35] Such reactions may be triggered by contact with air as soon as the sample is extracted from the arc chamber, starting at sample surface and being propagated to subsurface layers. The large concentration of O atoms measured in this region by EDS supports this mechanism, since the O atoms would not be only occupying adsorption sites on the sample surface, but they would be also substituting S atoms at the edges of $MoS_2$ subsurface layers.

From the Raman and SEM analysis exposed above, we can conclude that the final product of the pulsed arc process is a mixture of Mo, $MoS_2$ and C. The elemental and topographic surface characteristics suggest that the deposited layer is structured in: (1) dominant superficial population of $MoS_2$ grain fragments, (2) C-Mo thin layer as smoother background, and (3) near substrate deposition of micron-sized flakes of $MoS_2$ constituted of atomic few layers. The main mechanism responsible of $MoS_2$ flake formation, i.e. direct growth of $MoS_2$ islands on substrate surface or earlier production by fusing of precursor particles in plasma phase, supposes a fundamental question that would require the application of *in situ* and real time plasma process diagnostics.[36]

*3.2.4. High-resolution imaging and crystallinity (HRTEM, SAED, XRD)*

The crystallographic properties of the sample have been studied by powder XRD. Fig. 9a shows the diffractogram corresponding to $MoS_2$ powder precursor. The diffraction pattern corresponds to molybdenite, which crystallizes in the hexagonal P63/mmc group. On the other hand, Fig. 9b shows the XRD spectra corresponding to material deposited onto the collecting probe. In this case, only diffraction peaks corresponding to Tungsten Bragg reflections (bcc structure) are detected. The Tungsten signal is caused by the interaction of the X-rays with the W probe. The lack of XRD signal from $MoS_2$ or other materials suggest that the deposited layer is amorphous. Note that this conclusion is extracted from powder X-ray analysis that averages over a probing area of a few millimetres. Such probing size may neglect the contribution of eventual nanocrystals diluted within an amorphous matrix. In other words, such analysis is not suited for diagnosing samples containing highly diluted nanosized crystals, since their weak XRD signal could be overlapped with the spectrum background. The following paragraph discusses the localization and identification of nanocrystals by means of high-resolution TEM and SAED.

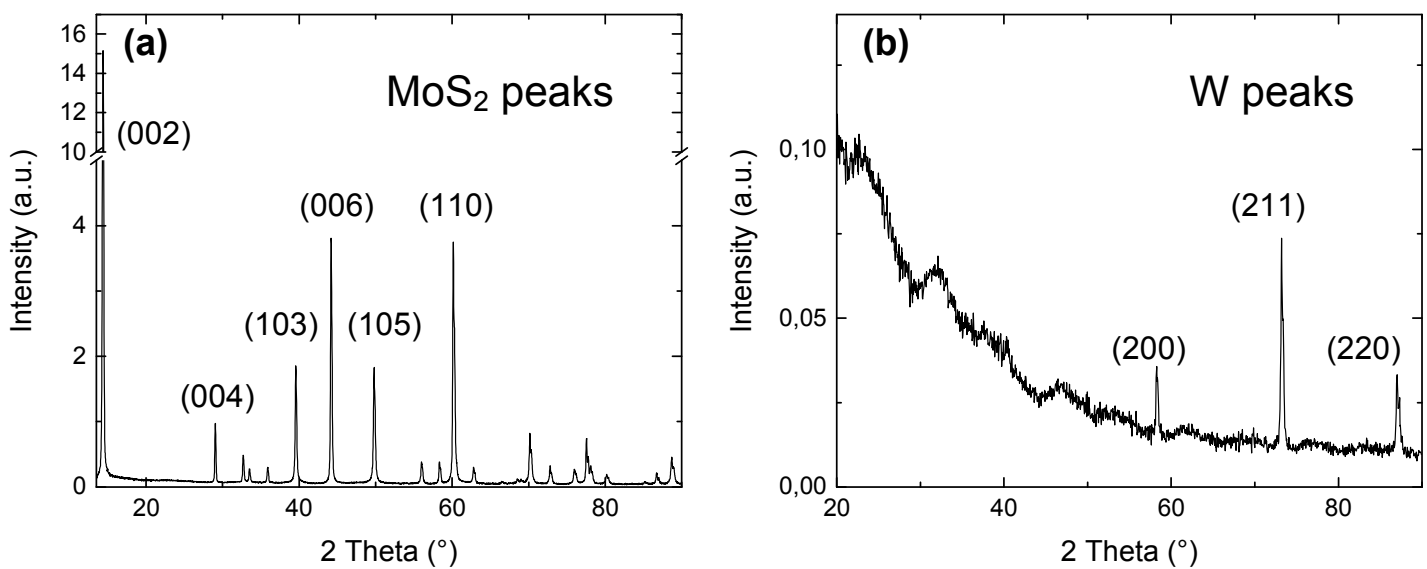

**FIG 9:** XRD diffractograms showing the crystalline structure of the samples. Most intense peaks are indexed. (a) $MoS_2$ powder precursor. (b) $MoS_2$ deposit on collecting probe measured at different positions showed similar spectra (measured intervals: 5 mm). The XRD spectra suggest an amorphous sample (or very diluted nanocrystals) along the deposition probe: only the W peaks are present in the spectra. The presence of nanocrystals is explored by HRTEM and SAED in Fig. 10.

Further study on sample crystallinity has been performed by TEM analysis (Fig. 10). A small amount of layer material from the middle region of the probe was detached and deposited onto a TEM copper grid. Also, a small sample of $MoS_2$ powder precursor was placed on a different Cu grid. Fig. 10a shows a representative high-resolution TEM image of $MoS_2$ powder precursor, which is compared with the HRTEM image of Fig. 10b, corresponding to the material sampled from the W probe. The $MoS_2$ grains exhibit large monocrystalline domains, as it is evidenced by the contrasted crystallographic planes visible in Fig. 10a. The accompanying dotted SAED pattern (inset) confirms this picture. On the other hand, the HRTEM image in Fig. 10b shows that the layer deposited on the probe during pulsed arc discharge is constituted by an amorphous phase combined with randomly oriented nanosized crystals. Additionally, the periodically modulated contrast intensities with elongated shapes in the image suggest the presence of carbon nanotubes combined with Moiré patterns due to interference effects of crystal lattice patterns. The polycrystalline phase present in the sample is manifested by the inset diffraction ring pattern, which reveals the presence of $MoS_2$ nanocrystals together with further material crystalline phases. In conclusion, we have observed that the material deposited onto the W probe is composed of nanosized $MoS_2$ crystallites embedded within a C-Mo amorphous matrix.

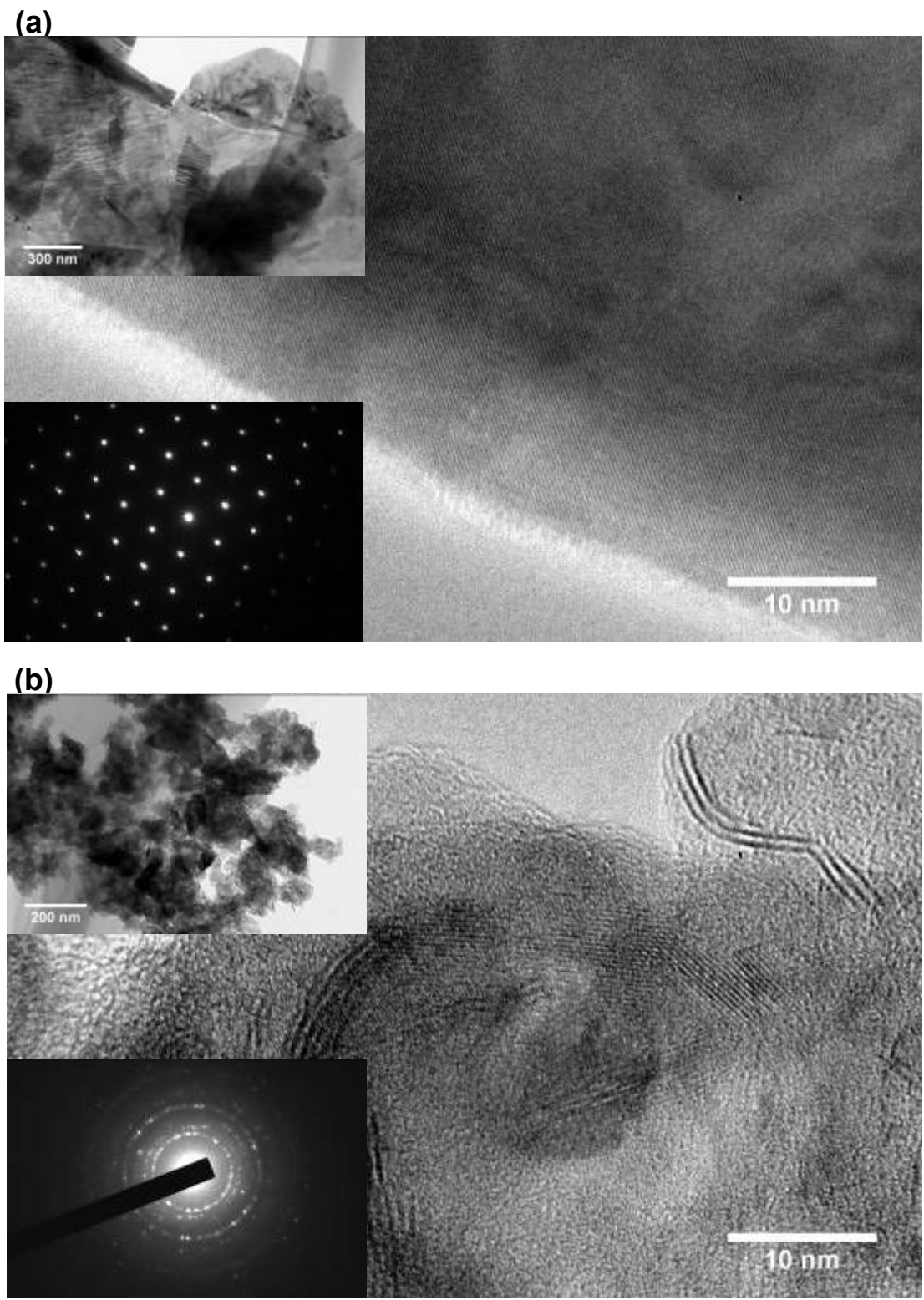

**FIG 10:** TEM micrographs of MoS$_2$ powder grain (a) and material detached from the wire sample (b). In (a), regions of the powder grains are characterized by large crystalline domains, as confirmed by the dotted SAED pattern. In (b), the presence of nanotubes is combined with the observation of Moiré patterns. The HRTEM images and ring-shaped SAED patterns indicate the presence of nanocrystals randomly oriented within the sample.

## 4. Conclusions

Few-layer synthesis of $MoS_2$ flakes has been successfully carried out by pulsed arc discharge from a compound anode. Such anode consisted of a hollow graphite anode filled with $MoS_2$ powder precursor. The synthesis has been possible using a non-reactive atmosphere ruling out the use of sulphur. Thus, this method requires only He gas injection and it is therefore environmentally friendly. The pulsed anodic arc discharge of the compound anode leads to a balanced evaporation of the filling powder and of the graphite anode, thus avoiding preferential ablation of the hollow graphite anode promoted in standard DC arc discharges. The shapes of the arc current and arc voltage waveforms were similar to those measured in pulsed carbon ablation from previous works. The electrical plasma parameters in $MoS_2$ arc discharge, such as electron density and ionic current, are of the same order of magnitude of the parameters characteristic of graphite arc discharge. The deposited material onto a collecting probe near the arc core consisted of fragments of $MoS_2$ powder, C and metallic Mo background, and few-layer flakes of $MoS_2$ attached to the probe surface. The obtained products demonstrate that the thermal energy provided by pulsed arc plasma is sufficient to dissociate molecules and, subsequently, to promote the required plasma chemistry to transform 3D $MoS_2$ material into 2D $MoS_2$ material on the substrate surface.

Future work towards understanding the growth mechanisms of Mo nanoparticles and $MoS_2$ flakes is planned. Advanced optical diagnostics would provide the necessary information to monitor particle dynamics in the arc plasma process. Also, an increase in purity of the deposited material, i.e. lower carbon contamination, is desirable and could likely be achieved by fine-tuning the shape of the excitation waveform signal. This objective will be fulfilled by employing a new power supply equipped with rapid switching circuit, which should render possible the operation of well-defined pulse signals yet in shorter duty cycles. Nevertheless, this study demonstrates that the use of pulsed power to excite an anodic arc discharge is a versatile technique to synthesize new nanomaterials. Besides, arc discharge operation in pulsed mode provides a more stable discharge dynamics and better control of arc processes compared to traditional DC discharges. The outcome of this study is encouraging in terms of finding new pathways of nanosynthesis by means of pulsed anodic arc discharges held near atmospheric pressure.

## Acknowledgments

This work was supported by the US Department of Energy, Office of Science, Fusion Energy Sciences program Award Number DESC0015767, and by the National Science Foundation Grant Number 1747760. The authors thank the support by Dr. Jiancun Rao from AIMLab at the Maryland NanoCenter. The assistance by Prof. Peter Y. Zavalij from the X-ray Crystallographic Center at University of Maryland is also acknowledged.